\documentclass{svjour3}                     
\smartqed  
\usepackage{graphicx}
\usepackage{amsmath}

 \usepackage{mathptmx}      
\usepackage{latexsym}
 \journalname{Few-Body Systems}
\begin{document}

\title{Treatment of confinement in the Faddeev approach to three-quark problems
}


\author{J. McEwen \and J. Day \and A. Gonzalez \and Z. Papp \and W. Plessas}
\institute{ J.\ McEwen 
\and
J.\ Day 
\and
A.\ Gonzalez 
\and
Z.\ Papp 
\at Department of Physics and Astronomy,
California State University Long Beach, Long Beach, California, USA 
\and
W. Plessas
\at Institute for Theoretical Physics,
University of Graz, Graz, Austria}



\date{ Preprint: CSULB-PA-10-01, Received: date / Accepted: date}

\maketitle

\begin{abstract}
A method is presented that allows to solve the Faddeev integral equations of the semirelativistic constituent quark 
model. In such a model the quark-quark interaction is modeled by a infinitely rising confining potential and the 
kinetic energy is taken in a relativistic form. We solve the integral equations in Coulomb-Sturmian basis. 
This basis facilitate an exact treatment of the confining potentials.

\end{abstract}

\section{Introduction}
\label{intro}

The Faddeev approach to quantum-mechanical three-body problems has been efficiently applied
to a variety of problems for many years. The main advantage over other methods consists of
solving general Hamiltonians with two and three-body forces, implementing a priori the proper boundary conditions 
(bound state and/or scattering), and in observing the proper symmetries for (non)identical particles. 
While problems with short-range forces can be treated 
in a straightforward manner, special care must be exerted if long-range interactions are present. 
This applies in particular to the three-body Coulomb problem, but also to the case when confining
interactions are present. The reason is that any long-range force influences the asymptotic
behavior such that the system never becomes free of interactions. As a result, in the
Coulomb case one has infinitely many bound states accumulating at zero energy, whereas for
the confinement there exist infinitely many bound states with arbitrarily immense binding energies.

Faddeev equations have already been applied in the past to confined
three-quark systems. The approach was first developed for a confinement of harmonic
oscillator type, and one has achieved an exact solution for the three-body ground
state~\cite{Friar:1980za}. The same method was subsequently followed for a
non-relativistic three-quark Hamiltonian, including
the addition of a linear confinement to a hyperfine interaction~\cite{SilvestreBrac:1985ic,Glozman:1996wq}.
The technique always consisted of defining the zero point of the spectrum such that the
three-body states in question become deeply bound and standard methods become applicable. 
One of the notable disadvantages of this approach
consists of difficulties in the numerical treatment as the partial-wave expansion becomes only
slowly convergent.

Some time ago, one of us proposed an alternative method for dealing with confinement in three-quark systems~\cite{Papp:1998yt}. It is similar to the one followed in the Coulomb case and consists of
splitting the quark-quark potential into a long-range confining part and a short-range (non-confining)
interaction. The Faddeev procedure is then applied only to the non-confining short-range parts. The resulting integral equations were solved by using the Coulomb-Sturmian separable expansion
approach. Subsequently the method was successfully applied to a semirelativistic constituent-quark
model~\cite{Papp:2000kp}.

In this approach one requires the evaluation of the matrix elements in the Green's operator
corresponding to the long-range part of the Hamiltonian between the basis states of the
separable expansion. Whereas in the Coulomb case it is readily possible to compute these
matrix elements in the Coulomb-Sturmian basis. An analogous result has not been known for the
confining interactions, rather the confinement has been treated in an approximate
way~\cite{Papp:1998yt,Papp:2000kp}. The comparison with results for baryon masses obtained
from variational calculations suggested that this might not be a bad approximation, but one
certainly desires to get rid of it, especially with regard to getting reliable wave functions for
baryon ground and excited states.

In this paper we present a method to solve the Faddeev equations for confining interactions
without resorting to any such approximation. The technique makes use of a recent result obtained
for the Green's operator of a Hamiltonian consisting of interactions expressed as polynomials of
the inter-particle separations~\cite{Kelbert:2007}. In the following Section we shortly outline
the Faddeev approach to three-body problems. Then, in Section \ref{sec3}, we explain the
procedure for modifying the Faddeev approach in order to treat confining interactions.
Subsequently we present the solution of the Faddeev equations through the Coulomb-Sturmian expansion method. We give the results for the nucleon and $\Delta$ ground and excited states,
in the case of the relativistic Goldstone-boson-exchange constituent quark
model~\cite{Glozman:1997ag}. 

\section{Faddeev approach to three-particle problems}
\label{sec2}
We consider a three-particle Hamiltonian 
\begin{equation}\label{hamilton}
H=H^{(0)}+v_{1}+v_{2}+v_{3},
\end{equation}
where $H^{(0)}$ is the kinetic-energy operator and $v_{\alpha}$, $\alpha=1,2,3$, are the 
mutual interactions of the particles. We represent it through the usual configuration-space
Jacobi coordinates: e.g.\ ${\vec x}_{1}$ is the coordinate between particles $2$ and $3$ and ${\vec y}_{1}$ is the coordinate between the center of mass of the pair $(2,3)$ and particle $1$. Let us assume for a moment that the interactions $v_{\alpha}$ are short-range potentials.

We want to solve the Schr\"odinger equation
\begin{equation}\label{schrodinger}
H|\Psi\rangle=(H^{(0)}+v_{1}+v_{2}+v_{3})|\Psi \rangle = E |\Psi\rangle~.
\end{equation} 
Introducing 
\begin{equation}
G^{(0)}(E)=(E-H^{(0)})^{-1}
\end{equation}
 and rearranging (\ref{schrodinger}),  we have
\begin{eqnarray}\label{fdec1}
|\Psi\rangle && =  G^{(0)}(E)(v_{1}+v_{2}+v_{3})|\Psi \rangle \nonumber \\
&& = G^{(0)}(E)v_{1}|\Psi \rangle + 
G^{(0)}(E) v_{2} |\Psi \rangle +G^{(0)}(E)v_{3}|\Psi \rangle.
\end{eqnarray} 
This means that the three-particle wave function $|\Psi\rangle$ naturally breaks down to three 
components 
\begin{equation}\label{fdec2}
|\Psi\rangle=|\psi_{1} \rangle + |\psi_{2} \rangle + |\psi_{3} \rangle~,
\end{equation} 
where 
\begin{equation}\label{fcomp}
|\psi_{\alpha}\rangle=G^{(0)}(E)v_{\alpha}|\Psi \rangle, \ \ \ \ \alpha=1,2,3,
\end{equation}
are the Faddeev components. The Faddeev components satisfy the set of equations, the Faddeev equations,
\begin{eqnarray}\label{faddeqdiff}
(E-H^{(0)}-v_{1})|\psi_{1} \rangle &=& v_{1}( |\psi_{2} \rangle + |\psi_{3} \rangle) \\
(E-H^{(0)}-v_{2})|\psi_{2} \rangle &=& v_{2}( |\psi_{1} \rangle + |\psi_{3} \rangle) \\
(E-H^{(0)}-v_{3})|\psi_{3} \rangle &=& v_{3}( |\psi_{1} \rangle + |\psi_{2} \rangle)~. 
\end{eqnarray} 
By adding up these equations and considering (\ref{fdec1}) we recover the Schr\"odinger
equation. So, the Faddeev procedure is no more and no less than a way of solving the Schr\"odinger equation.
With the help of channel Green's operators 
\begin{equation}
G_{\alpha}(E)=(E-H^{(0)}-v_{\alpha})^{-1},
\end{equation}
 we can also recast the Faddeev equations into an integral equation form
\begin{eqnarray}\label{faddeqint}
|\psi_{1} \rangle &=& G_{1}(E) \: v_{1}( |\psi_{2} \rangle + |\psi_{3} \rangle) \\
|\psi_{2} \rangle &=& G_{2}(E)\: v_{2}( |\psi_{1} \rangle + |\psi_{3} \rangle) \\
|\psi_{3} \rangle &=& G_{3}(E)\: v_{3}( |\psi_{1} \rangle + |\psi_{2} \rangle)~. 
\end{eqnarray} 
  
 In general, the potentials may support bound states.  So, we may have a situation when particles $2$ and $3$ form a bound state and particle $1$
is far away. Alternatively particles $1$ and $2$ form a bound state and particle $3$ is away, and so on. So, in the three-particle wave function we may have three genuinely different two-body asymptotic channels. Although these two-body channels represent genuinely different physical situations, yet in solving the Schr\"odinger equation, we need to impose all these different boundary conditions on a single wave function. 

The Faddeev components posses a simpler structure. For example, in Eq.\ (\ref{fcomp}), the short range potential $v_{1}$ acting on $|\Psi\rangle$, suppresses those asymptotic structures when particles $2$ and $3$ are far away, i.e.\  it eliminates the two-body channels when either particle $2$ or particle $3$ is at infinity.
Consequently,  $|\psi_{1}\rangle$ contains only one kind of asymptotic channels, channels when particle $1$ is at infinity. A similar statement is valid for 
$|\psi_{2}\rangle$ and $|\psi_{3}\rangle$. So, with the Faddeev decomposition we achieve an asymptotic filtering, 
i.e.\ we split the wave function into parts such that each component possess only one kind of asymptotic behavior. Then solving the Faddeev differential equation we need to impose simpler conditions on the components.

\section{Faddeev equations for confining potentials}\label{sec3}

It is obvious that if we apply the Faddeev decomposition to long-range confining potentials, we run into trouble.
If $v_{\alpha}$ in Eq.\ (\ref{fcomp}) is of long-range type, then applying on $|\Psi\rangle$, it would not suppress the components where particles $\beta$ or $\gamma$ are far away. The  procedure does not lead to any simplification in the asymptotic behavior of the components: the whole Faddeev procedure does not make much sense.

A similar situation occurred when the Coulomb potential was directly inserted into the Faddeev equations. This approach turned out to be a complete failure. The way out of the trouble was proposed by Merkuriev \cite{merkuriev1980three}. He proposed to split the Coulomb potential into long-range and short-range parts 
and applied the Faddeev procedure only to the short range parts. The set of Faddeev-Merkuriev integral equations possesses a compact kernel and its asymptotic analysis provided the boundary conditions to the differential equations \cite{merkuriev1980three}. The three-particle systems with attractive Coulomb potentials became amenable to numerical treatment both in differential \cite{PhysRevLett.33.1350,PhysRevA.45.2723} and integral equation formalism \cite{PhysRevA.63.062721}.

Inspired by Merkuriev's treatment of the Coulomb potential we adopted \cite{Papp:1998yt} a separation of the 
confining potential into long range confining and short range non-confining terms
\begin{equation}\label{vsplit}
v_{\alpha}=v_{\alpha}^{(c)}+v_{\alpha}^{(s)},
\end{equation}
where $v_{\alpha}^{(c)}$ and $v_{\alpha}^{(s)}$ are confining and short range potentials, respectively.
Then we can write the Hamiltonian (\ref{hamilton}) in the form
\begin{equation}\label{h}
H=H^{(c)}+v_{1}^{(s)}+v_{2}^{(s)}+v_{3}^{(s)}~,
\end{equation}
where 
\begin{equation}\label{hc}
H^{(c)}=H^{(0)}+v_{1}^{(c)}+v_{2}^{(c)}+v_{3}^{(c)}~.
\end{equation}  

Eq.\ (\ref{h}) looks like an ordinary Hamiltonian with short-range potentials. Therefore the Faddeev procedure is readily applicable and the Faddeev components are defined by
\begin{equation}\label{fcompc}
|\psi_{\alpha}\rangle=G^{(c)}(E)v_{\alpha}^{(s)} |\Psi \rangle, \ \ \ \ \alpha=1,2,3,
\end{equation}
where $G^{(c)}(E)=(E-H^{(c)})^{-1}$. Since $v_{\alpha}^{(s)}$ is of short-range type, its acting on $|\Psi\rangle$ 
suppresses those asymptotic parts in $|\Psi\rangle$ which are related to situations where particles
$\beta$ and $\gamma$ are far away. With this definition of the asymptotic components,  the Faddeev integral equations take the form
\begin{eqnarray}\label{faddeqs}
|\psi_{1} \rangle &=& G_{1}^{(c)}(E)\: v_{1}^{(s)}( |\psi_{2} \rangle + |\psi_{3} \rangle) \\
|\psi_{2} \rangle &=& G_{2}^{(c)}(E)\: v_{2}^{(s)}( |\psi_{1} \rangle + |\psi_{3} \rangle) \\
|\psi_{3} \rangle &=& G_{3}^{(c)}(E)\: v_{3}^{(s)}( |\psi_{1} \rangle + |\psi_{2} \rangle)~,
\end{eqnarray} 
where
\begin{equation}
G^{(c)}_{\alpha}(E)=(E-H^{(c)}-v_{\alpha}^{(s)})^{-1}~.
\end{equation}

With this treatment of the confining potential we achieved that the Faddeev decomposition acts as 
an asymptotic filtering, and the Faddeev integral equations formally contains only short-range potentials. 
All the long range terms are kept in the Green's operators. However, we should be careful here. The 
Hamiltonian $H^{(c)}$ contains the long-range confining parts of the original Hamiltonian. Thus $H^{(c)}$ has infinitely many discrete eigenstates. At those discrete energies 
$G^{(c)}(E)$ is singular. So, acting on a vanishing $v_{\alpha}^{(s)}|\Psi\rangle$, they may produce a 
non-vanishing $|\psi_{\alpha}\rangle$. Thus it may happen that at certain energy the 
Faddeev equations have solutions where none of the $|\psi_{\alpha}\rangle$ components are 
zero but their sum, the total wave function $|\Psi\rangle$, vanishes. 
These solutions are spurious or ghost solutions. They are 
due to the pole of $G^{(c)}(E)$. However, $G^{(c)}(E)$ is just an auxiliary operator. It depends on the splitting
of $v_{\alpha}$. Therefore the spurious solutions depends on the splitting, while the real solutions should not.
If we vary the way we split the potential in Eq.\ (\ref{vsplit}) we
can shift the energy of the spurious solutions and with a careful choice of $v^{(c)}$ 
we can push them out of the energy region of physical interest.

A further advantage of the Faddeev method that the identity of particles greatly simplifies the equations. If particles $1$, $2$, and $3$ are identical,
then $\psi_{1}$, in its natural Jacobi coordinate system $\{ {\vec x}_{1}, {\vec y}_{1}\}$, looks like $\psi_{2}$ in its natural Jacobi coordinate system $\{{\vec x}_{2}, {\vec y}_{2}\}$ and $\psi_{3}$ in its natural Jacobi coordinate system $\{{\vec x}_{3},{\vec y}_{3}\}$.
On the other hand, the wave function is either symmetric ($p=1$) or antisymmetric ($p=-1$) with respect to interchanging 
two particles
\begin{equation}
{\mathcal P}_{23} | \psi_1 \rangle = p | \psi_1 \rangle.
\end{equation}
Building this information into the formalism we arrive at
a single integral equation
\begin{equation} \label{fmp}
| \psi_{1} \rangle =  2 G_1^{(c)} v_1^{(s)} {\mathcal P}_{123}
| \psi_{1} \rangle,
\end{equation}
where ${\mathcal P}_{123}={\mathcal P}_{12}{\mathcal P}_{23}$ is the operator for cyclic permutation of all three particles 
${\mathcal P}_{123} |\psi_{1}\rangle =| \psi_{2}\rangle$. The three-quark system is completely
antisymmetric in terms of color quantum numbers, so it should be symmetric in the other quantum 
numbers. To achieve the proper symmetry for all the three quarks we only need to ensure a symmetric 
wave function with respect to exchange only two particles. This can easily be achieved by selecting the
subsystem angular momentum $l$, spin $s$ and isospin $t$ such that
\begin{equation}\label{symcond}
(-)^{l+s+t}=1.
\end{equation}
This will ensure the correct symmetry for all the three quarks in the baryon.

\section{Solution of the Faddeev equations for confining potentials}
\label{sec4}

We solve the Faddeev equations be using the Coulomb-Sturmian (CS) separable expansion approach.
In configuration space the CS functions are defined by
\begin{equation}
\langle r | n l ; b\rangle = \sqrt{ \frac{ n!}{(n+2l+1)!} } (2br)^{l+1} \exp(-br) \: L_{n}^{2l+1}(2br),
\end{equation}
where $n=0,1\ldots$,  $l$ is the angular momentum, $b$ is a parameter and $L$ is the Laguerre polynomial. The biorthonormal partner to $|nl;b\rangle$ is defined as $\langle r | \widetilde{nl;b} \rangle = 1/r  \cdot \langle r | nl;b \rangle$. The orthogonality and completeness, in angular momentum subspace, 
is given by
\begin{equation}
\langle n' l ;b| \widetilde{n l;b} \rangle = \langle  \widetilde{n' l;b } | n l;b \rangle =\delta_{n' n}
\end{equation}   
and
\begin{equation}
\lim_{N\to\infty} \sum_{n=0}^{N} | \widetilde{n l;b} \rangle   \langle n l ;b| = \lim_{N\to\infty} \sum_{n=0}^{N} |  n l;b \rangle  \langle \widetilde{nl;b} | = \lim_{N\to\infty} {\bf 1}^{N}_{l}.
\end{equation}
The CS functions also have a nice analytic form in momentum space
\begin{equation}
\langle p | n l;b\rangle =  \sqrt{\frac{2}{\pi} \frac{n!}{(n+2l+1)!}} \frac{(n+l+1) l! b (4bp)^{l+1}}{(p^{2}+b^{2})^{2l+2}} \; G_{n}^{l+1}\left( \frac{p^{2}-b^{2}}{p^{2}+b^{2}}\right)
\end{equation}
and
\begin{equation}
\langle p | \widetilde{nl;b} \rangle = \frac{p^{2}+b^{2}}{2 b (n+l+1)}\langle p | n l;b\rangle.
\end{equation}

The three-body Hilbert space is the direct sum of two-body Hilbert spaces associated with coordinates $\vec x_{\alpha}$ and $\vec y_{\alpha}$. Then the appropriate basis is given by
\begin{equation}
| n l \nu \lambda \rangle_{\alpha} =\left\{ |nl;b_{x}\rangle_{\alpha}  \otimes |\nu \lambda ; b_{y}
\rangle_{\alpha} \right\},
\end{equation}
where $ |nl;b_{x}\rangle_{\alpha}$ and $|\nu \lambda;b_{y}\rangle_{\alpha}$ are associated with coordinates $\vec x_{\alpha}$ and $\vec y_{\alpha}$, respectively. We can have three different bases of this kind for $\alpha=1,2,3$, and the completeness relation takes the form
\begin{equation}
\begin{split}
{\bf 1} &=  \lim_{N\to\infty} \sum_{n,\nu}^{N} | \widetilde{n l \nu \lambda} \rangle_{\alpha} \; _{\alpha}\langle n l \nu \lambda | \\
& =\lim_{N\to\infty} \sum_{n,\nu}^{N} | {n l \nu \lambda } \rangle_{\alpha} \; _{\alpha}\langle \widetilde{n l \nu \lambda} | =\lim_{N\to\infty} {\bf 1}^{N}_{\alpha}~.
\end{split}
\end{equation}

In order to have a convenient representation for the Faddeev equations, we plug the unity operators into the equations
\begin{equation}\label{faddeqsapp}
|\psi_{\alpha} \rangle = \lim_{N\to\infty} G_{\alpha}^{(c)}(E) \left[ {\bf 1}^{N}_{\alpha} v_{\alpha}^{(s)}  {\bf 1}^{N}_{\beta} |\psi_{\beta} \rangle +  {\bf 1}^{N}_{\alpha} v_{\alpha}^{(s)} {\bf 1}^{N}_{\gamma}  |\psi_{\gamma} \rangle  \right]~.
\end{equation} 
This representation becomes an approximation if we keep $N$ finite. This is equivalent of approximating the short-range potentials in the three-body Hilbert space basis
\begin{equation}\label{valpha-approx}
v_{\alpha}^{(s)}\approx \sum_{n,\nu,n'\nu'}^{N} | \widetilde{n l \nu\lambda } \rangle_{\alpha} \;
 \underline{v}^{(s)}_{\alpha \beta}\;_{\beta}\langle \widetilde{n' l' \nu' \lambda'} | ~,   
\end{equation}
where
$ \underline{v}^{(s)}_{\alpha \beta} = \mbox{}_{\alpha}\langle n \nu | v_{\alpha}^{(s)} |{n' \nu'} \rangle_{\beta}$. This approximation turns the Faddeev equations into a matrix equation
\begin{eqnarray}\label{feqmat}
\underline{\psi}_{\alpha}  &=& \underline{G}_{\alpha}^{(c)}(E) \:(  \underline{v}_{\alpha \beta}^{(s)} \underline{\psi}_{\beta}  +\underline{v}_{\alpha \gamma}^{(s)}  \underline{\psi}_{\gamma} ) \\
\end{eqnarray} 
where
\begin{equation}
\underline{G}^{(c)}_{\alpha}(E)= \mbox{}_{\alpha}\langle \widetilde{n l \nu \lambda }|(E-H^{(c)}-v_{\alpha}^{(s)})^{-1}| \widetilde{n' l' \nu' \lambda' }\rangle_{\alpha}~.
\end{equation}
By rearranging, we get a homogeneous linear algebraic equation for the component vector 
\begin{equation}
(\underline{{\cal G}}^{-1} -\underline{{\cal V}})\underline{\psi} = 0,
\end{equation}
where
\begin{equation}
\underline{{\cal G}}^{-1}=\begin{pmatrix}
(\underline{G}^{(c)}_{1})^{-1}(E) & \underline{0} &  \underline{0}    \\
\underline{0} & (\underline{G}^{(c)}_{2})^{-1}(E) & \underline{0}    \\
\underline{0} &  \underline{0} & (\underline{G}^{(c)}_{3})^{-1}(E)  
\end{pmatrix},
\end{equation}
\begin{equation}
\underline{{\cal V}}=\begin{pmatrix}
 \underline{0} &  \underline{v}^{(s)}_{12}  &  \underline{v}^{(s)}_{13}   \\
 \underline{v}^{(s)}_{21}  &  \underline{0} &  \underline{v}^{(s)}_{23}     \\
 \underline{v}^{(s)}_{31}  &  \underline{v}^{(s)}_{32}  & \underline{0}
\end{pmatrix},
\end{equation}
and
\begin{equation}
\underline{\psi}=\begin{pmatrix}
 \underline{\psi}_{1}  \\
 \underline{\psi}_{2}   \\
 \underline{\psi}_{3}  
\end{pmatrix}.
\end{equation}
This equation is solvable if and only if the determinant vanishes
\begin{equation}
\det (\underline{{\cal G}}^{-1} -\underline{{\cal V}}) ={0}.
\end{equation}
While the numerical evaluation of the matrix elements $\underline{v}_{\alpha \beta}^{(s)}$ is straightforward by using the transformation of Jacobi coordinates, 
to determine the matrix elements $\underline{G}^{(c)}_{\alpha}$ we need further approximations.

\subsection{The CS matrix elements of $G^{(c)}_{\alpha}$}
\label{sec:2.1}

The Green's operator $G^{(c)}_{\alpha}$ is the resolvent of a complicated three-body operator
\begin{equation}
H^{(c)}_{\alpha}=H^{(0)}_{rel} + \sum_{i=1}^{3} v_{i}^{(c)} + v_{\alpha}^{(s)}=H^{(0)}_{rel} +  v_{\beta}^{(c)} +  v_{\gamma}^{(c)}+ v_{\alpha}~.
\end{equation}
Here
\begin{equation}\label{relkin}
H^{(0)}_{rel}=\sum_{i=1}^{3} \sqrt{k_{i}^{2} +m_{i}^{2} }
\end{equation}
is the relativistic kinetic energy operator,
$m_{i}$ is the mass of the constituent quarks  and $\vec{k}_{i}$ are the individual quark three-momenta in the
frame where the total three-momentum $\vec{P}=\sum_{i=1}^{3} \vec{k}_{i}=0$.
Since this operator is too complicated for a straightforward evaluation of its resolvent we adopt the following procedure. We separate off a Hamiltonian which describes the asymptotically most relevant part of 
$H_{\alpha}^{(c)}$, and which we try to handle without approximations. The rest, which is not dominant asymptotically, can be approximated on the CS basis.

We define the operator $\widetilde{H}_{\alpha}$ as a sum of two-body operators acting on $x_{\alpha}$ and $y_{\alpha}$, respectively,
\begin{equation}
\widetilde{H}_{\alpha}=h_{x_{\alpha}}(x_{\alpha})+h_{y_{\alpha}}(y_{\alpha}),
\end{equation}
where
\begin{equation}
h_{x_{\alpha}}(x_{\alpha})= \sqrt{p_{x_{\alpha}}^{2}+\mu_{x_{\alpha}}^{2} } +v_{\alpha}(x_{\alpha})
\end{equation}
and 
\begin{equation}
h_{y_{\alpha}}(y_{\alpha})= \sqrt{p_{y_{\alpha}}^{2}+\mu_{y_{\alpha}}^{2} } +u^{(c)}_{\alpha}(y_{\alpha}).
\end{equation}
Here  $\mu_{x_{\alpha}}=m_{\beta} m_{\gamma}/(m_{\beta} + m_{\gamma} )$
and
$
\mu_{y_{\alpha}}= m_{\alpha}(m_{\beta} +m_{\gamma})/(m_{\alpha} +m_{\beta} + m_{\gamma} )
$ are the reduced masses.
The operator $h_{x_{\alpha}}$ is the semi-relativistic Hamiltonian of the $(\beta,\gamma)$ pair, while
the operator $h_{y_{\alpha}}$ describes the asymptotic motion
of the particle $\alpha$ and the pair $(\beta,\gamma)$. 
The auxiliary confining potential $u^{(c)}_{\alpha}(y_{\alpha})$ should behave
as $v_{\beta}^{c}(x_{\beta}) +  v_{\gamma}^{c}(x_{\gamma})$ for large $y_{\alpha}$ distances.

 The Green's operator $G^{(c)}_{\alpha}$ satisfies
 the resolvent relation
\begin{equation}\label{gcl}
G^{(c)}_{\alpha}=\widetilde{G}_{\alpha} + \widetilde{G}_{\alpha} U_{\alpha}G^{(c)}_{\alpha}~, 
\end{equation}
where $\widetilde{G}_{\alpha}(z)=(z-\widetilde{H}_{\alpha})^{-1}$ is the resolvent of $\widetilde{H}_{\alpha}$ and
\begin{equation}
\begin{split}
U_{\alpha}=&H^{(c)}_{\alpha}-\widetilde{H}_{\alpha} \\
=&\sum_{i=1}^{3} \sqrt{k_{i}^{2}+m_{i}^{2}} - \sqrt{p_{x_{\alpha}}^{2}+\mu_{x_{\alpha}}^{2} } -\sqrt{p_{y_{\alpha}}^{2}+\mu_{y_{\alpha}}^{2} } +v_{\beta}^{(c)}(x_{\beta})+v_{\gamma}^{(c)}(x_{\gamma}) - u^{(c)}_{\alpha}(y_{\alpha})~.
\end{split}
\end{equation}

In calculating the CS matrix elements of $G^{(c)}_{\alpha}$ we again make the separable approximation in Eq.\ (\ref{gcl}) and get
\begin{equation}\label{ualpha-approx}
\left(\underline{G}^{(c)}_{\alpha}(E)\right)^{-1} = \left(\underline{\widetilde{G}}_{\alpha}(E)\right)^{-1} -\underline{U}_{\alpha}~,
\end{equation}
where $\underline{\widetilde{G}}_{\alpha}(E)=\mbox{}_{\alpha}\langle \widetilde{ n \nu }| \widetilde{G}_{\alpha}(E) | \widetilde{n' \nu'} \rangle_{\alpha}$ and
$\underline{U}_{\alpha}=\mbox{}_{\alpha}\langle { n \nu }| U_{\alpha} | {n' \nu'} \rangle_{\alpha}$. 
The later matrix elements can also be evaluated numerically partly in configuration, 
partly in momentum space.

\subsection{The matrix elements $\underline{\widetilde{G}}_{\alpha}$} 

For calculating the matrix elements of ${\widetilde{G}}_{\alpha}$  we can adopt the technique used 
before in the Coulomb case  \cite{Papp:1996yf}. The method is based on the Riesz-Dunford 
functional calculus. Given a self-adjoint operator $h$ on a Hilbert space and an analytic function $f$ in 
the neighborhood of $\sigma(h)$, the spectrum of $h$, the function of the operator 
$f(h)$ is defined by a Cauchy-type contour integral
\begin{equation}
f(h)=\frac{1}{2\pi i} \oint_{\cal C} f(z) (z-h)^{-1} dz~,
\end{equation}
where $\cal C$ the boundary of a domain $D \supset \sigma(h)$.

The Green's operator $\widetilde{G}_{\alpha}$ is a function of the self-adjoint operator $h_{x_{\alpha}}$.
Therefore
\begin{equation}\label{gtilda}
\begin{split}
\widetilde{G}_{\alpha}(E)&=(E-h_{y_{\alpha}}-h_{x_{\alpha}})^{-1} \\
&= \frac{1}{2\pi i} \oint_{\cal C} (E-h_{y_{\alpha}}-z)^{-1}  (z-h_{x_{\alpha}})^{-1} dz \\
&= \frac{1}{2\pi i} \oint_{\cal C} g_{y_{\alpha}}(E-z) \: g_{x_{\alpha}}(z) \:dz~,
\end{split}\end{equation}
where $g_{y_{\alpha}}(z)=(z-h_{y_{\alpha}})^{-1}$ and $g_{x_{\alpha}}(z)=(z-h_{x_{\alpha}})^{-1}$. 
The contour $\cal C$ should be taken such that it encircles the singularities of $g_{x_{\alpha}}$ and 
avoids the singularities of $g_{y_{\alpha}}$.

\begin{figure}
\includegraphics[width=8cm]{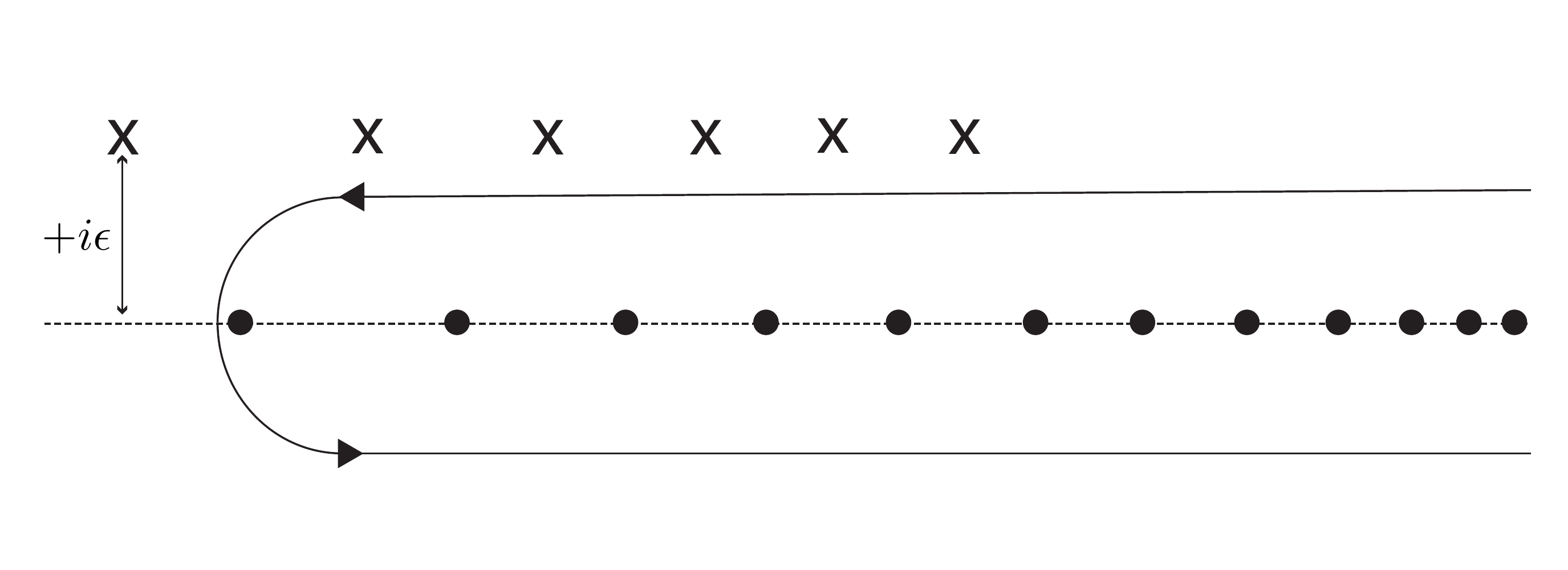}
\caption{Analytic structure of the integrand in Eq.\ (\ref{gtilda}) for complex 
$E=E_{r}+\mathrm{i} \epsilon$. The contour $\cal C$ circumvents the poles of $g_{x_{\alpha}}$ 
without incorporating the poles of $g_{y_{\alpha}}$. }
\label{fig:1}      
\end{figure}

\begin{figure}
\includegraphics[width=8cm]{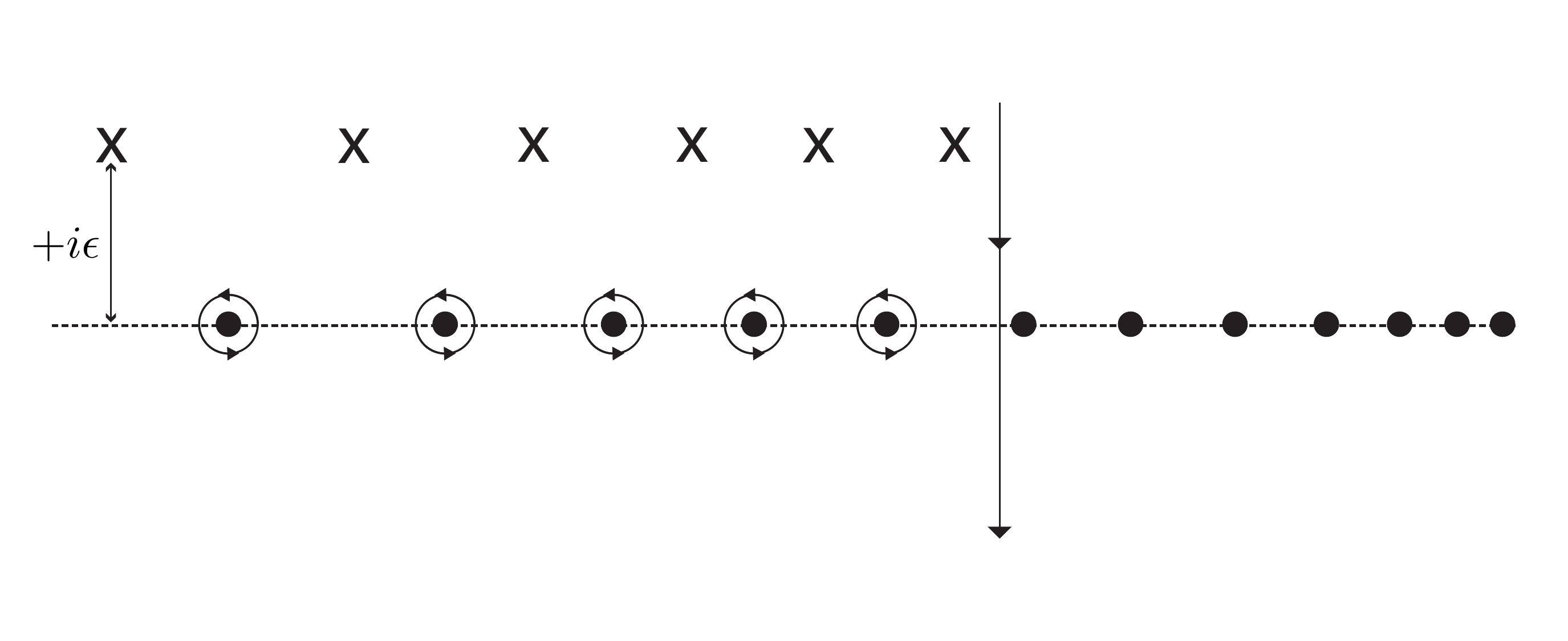}
\caption{The contour in Figure \ref{fig:1} is modified such that it shrinks to a few low-lying poles of 
$g_{x_{\alpha}}$ and opens up to a contour along imaginary direction. }
\label{fig:2}      
\end{figure}

\begin{figure}
\includegraphics[width=8cm]{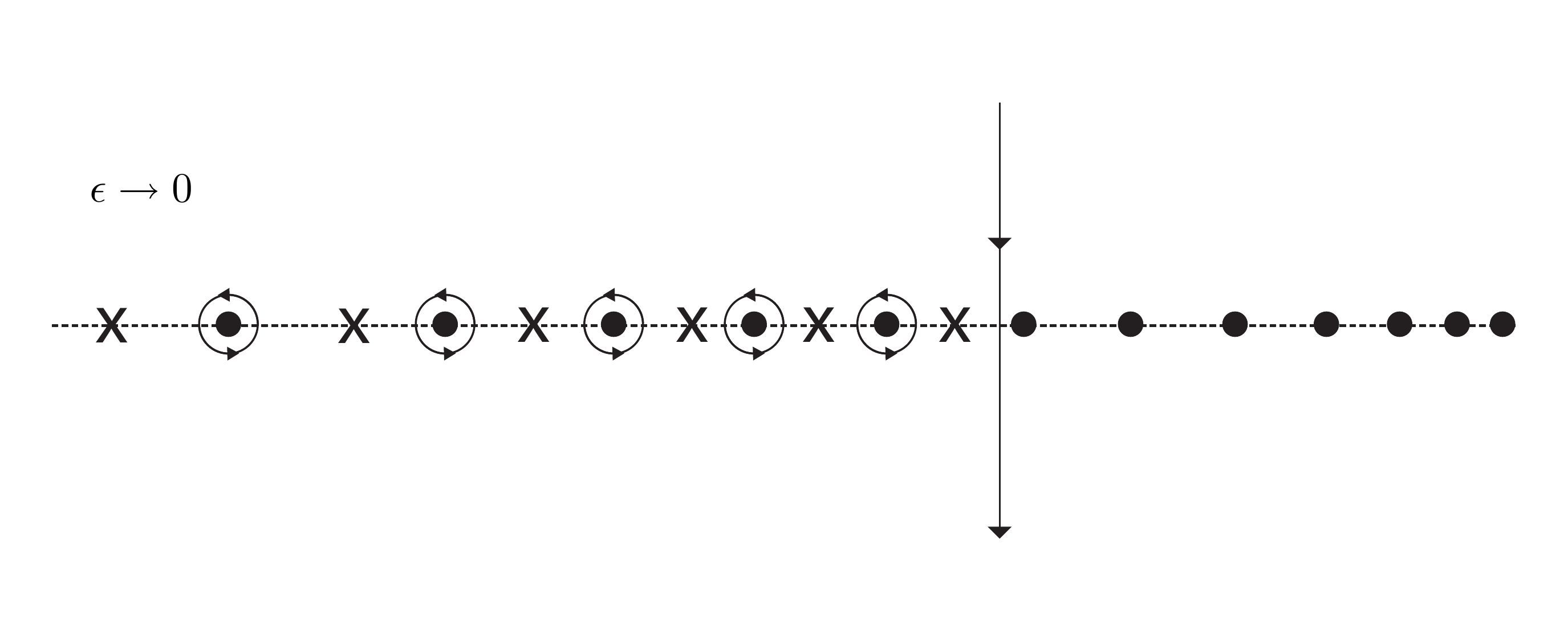}
\caption{ Now the contour avoids the singularities of $g_{y_{\alpha}}$ even in the $\epsilon \to 0$ limit.}
\label{fig:3}       
\end{figure}

In fact, both $h_{x_{\alpha}}$ and $h_{y_{\alpha}}$ are two-body Hamiltonians with confining potentials 
having discrete spectra. Figure \ref{fig:1} shows the analytic structure of the integrand in 
Eq.\ (\ref{gtilda}). For the better visibility, we display the integrand for complex energy 
$E=E_{r}+\mathrm{i}\epsilon$, $\epsilon  > 0$. 
Then the poles of $g_{y_{\alpha}}$ become well separated and we can easily draw a contour 
$\cal C$ that encircles the poles of $g_{x_{\alpha}}$ and avoids the poles of $g_{y_{\alpha}}$. 
In Figure \ref{fig:2} the contour $\cal C$ is deformed such that it shrinks to a few lowest-lying poles of 
$g_{x_{\alpha}}$ and opens up to an integral along imaginary direction. 
Now, even in the $\epsilon\to 0$ limit, the mathematical requirements for the contour integral are 
satisfied (Figure \ref{fig:3}). 

\subsection{Matrix elements of two-body Green's operators}

To evaluate $\underline{\widetilde{G}}_{\alpha}$ by a contour integral of two 
body Green's operators, we need the CS matrix elements of 
$g_{x_{\alpha}}$ and $g_{y_{\alpha}}$. Both $h_{x_{\alpha}}$ and $h_{y_{\alpha}}$ are 
semirelativistic Hamiltonians with short-range term
\begin{equation}\label{hsemi}
h=\sqrt{p^{2}+m^{2}}+a_{1}r+v^{(s)}.
\end{equation}

We can approximate the short-range potential $v^{(s)}$ in a separable way using a 
double CS basis \cite{Darai:2001yj}.  It amounts to expanding $v^{(s)}$ as
\begin{eqnarray}
v^{(s)} &\approx & \sum_{n m m' n'}^{N} | \widetilde{ nl; b}  
\rangle \left( \langle  \widetilde{nl;b}| ml;b' \rangle \right)^{-1} \langle m l ; b' | v^{(s)} | m' l'; b' \rangle  \left( \langle   m'l';b'  | \widetilde{n'l';b} \rangle \right)^{-1} 
\langle \widetilde{ n'l'; b} |  ~ \nonumber \\
&\approx & \sum_{n n'}^{N} | \widetilde{ nl; b}  \rangle 
\underline{v}^{(s)} 
\langle \widetilde{ n'l'; b} | ~.
\label{vsep}
\end{eqnarray}
This is  an exact representation if $N$ goes to infinity and becomes an approximation if $N$ is kept finite. 
The CS matrix elements of the potential have to be evaluated numerically.  This can easily be done  in configuration or momentum space, depending how the potential is defined.

For the operator $g(z)=(z-h)^{-1}$, we can immediately write down a resolvent equation
\begin{equation}\label{lsg}
g(z)=g^{(c)}(z)+g^{(c)}(z)v^{(s)}g(z),
\end{equation}
where $g^{(c)}(z)=(z-h^{(c)})^{-1}$, and 
\begin{equation}
h^{(c)}=\sqrt{p^{2}+m^{2}}+a_{1}r~.
\end{equation}
With the help of the approximation (\ref{vsep}) we can solve equation (\ref{lsg}), and get
\begin{equation}
\underline{g}(z)=((\underline{g}^{(c)}(z))^{-1}-\underline{v}^{(s)})^{-1}~.
\end{equation}
What remains is the evaluation of $\underline{g}^{(c)}(z)$. 

The main advantage of using CS functions in few-body Coulombic calculations comes from the fact that the Coulomb Green's operator can be given in analytic form. This is possible because in CS basis the Coulomb Hamiltonian is an infinite symmetric tridiagonal, or Jacobi, matrix. 
We found that the resolvent of a Hamiltonian having a Jacobi form can be given in terms of a continued fraction \cite{kelbert2007green}. 
This result can be generalized to polynomial potentials \cite{Kelbert:2007}. 
A non-relativistic kinetic energy with a polynomial potential has an infinite symmetric band structure. If the polynomial potential is a linear potential, 
the band matrix is penta-diagonal, if the confinement is quadratic, the band matrix is septa-diagonal. However, band matrices can be considered as tridiagonal 
matrices of block-matrices. In particular, a penta-diagonal band matrix can be considered as a tridiagonal matrix of $2\times 2$ block matrices. 

Unfortunately, the relativistic kinetic energy operator does not have a band structure on CS basis. But numerical studies revealed that for large $n$ and $n'$ values the dominant elements form a penta-diagonal band matrix. 
So, by neglecting the asymptotically not dominant elements 
outside the penta-diagonal band, we can use the matrix continued fraction method \cite{Day:2009ax}
\begin{equation}
\underline{g}^{(c)}(z) = (\underline{J} - \delta_{i, n'} \delta_{j, n'} J_{n',n'+1} C_{n'+1} 
J_{n'+1,n'})^{-1},
\end{equation}
where $C$ is a matrix continued fraction
defined recursively by
\begin{equation}\label{mconf}
C_{i+1}=(J_{i+1,i+1}-J_{i+1,i+2} C_{i+2} J_{i+2,i+1})^{-1}~
\end{equation}
where $J=z-h$ in CS representation, $J_{i,j}$ refers to $2\times2$ blocks and $n'=n/2$.
So,  $\underline{g}^{c}$ is an inverse of the modified $\underline{J}$ matrix. 
The modification is a matrix continued fraction $C$, and it only effects the bottom right  $2\times2$ block of $\underline{J}$. 
Numerical studies showed, that although we neglected the asymptotically non-dominant terms in $J$, this construction of $\underline{g}^{(c)}$ is very accurate  \cite{Day:2009ax}.

\section{Goldstone boson exchange model for baryons}\label{sec-gbe}

In this work we consider a semirelativistic quark model with Goldstone boson exchange quark-quark interaction. The kinetic energy operator is given in the form of Eq.\ (\ref{relkin}).
 The quark-quark interaction is derived from Goldstone-boson exchange model 
\begin{equation}
v_{\alpha}=V^{conf}_{\alpha}+V^{\chi}_{\alpha}~,
\end{equation}
where confinement potential is taken in the form
\begin{equation}
V^{conf}_{\alpha}=V_{0}+C x_{\alpha}~.
\end{equation}
The chiral potential is a sum of octet
\begin{equation}
V^{\chi}_{\alpha}(\vec{x}_{\alpha}) =
\sum_{F=1}^{3} V_{\pi} ({\vec x}_{\alpha})  \lambda_{\beta}^{F} \lambda_{\gamma}^{F} {\vec \sigma}_{\beta} {\vec \sigma}_{\gamma} +
\sum_{F=4}^{7} V_{K} ({\vec x}_{\alpha})  \lambda_{\beta}^{F} \lambda_{\gamma}^{F} {\vec \sigma}_{\beta} {\vec \sigma}_{\gamma} +
 V_{\eta} ({\vec x}_{\alpha})  \lambda_{\beta}^{8} \lambda_{\gamma}^{8}   
 {\vec \sigma}_{\beta} {\vec \sigma}_{\gamma},
\end{equation}
and singlet
\begin{equation}
V^{\chi}_{\alpha}(\vec{x}_{\alpha}) = 
 \frac{2}{3} V_{\eta'} ({\vec x}_{\alpha})   
 {\vec \sigma}_{\beta} {\vec \sigma}_{\gamma},
\end{equation}
 meson-exchange terms, where ${\vec \sigma}$ and $\lambda$ are the quark spin and flavor matrices, 
 $V_{\pi}$, $V_{K}$, $V_{\eta}$ and $V_{\eta'}$ represent the form factor
for $\pi$, $K$, $\eta$ and $\eta'$ meson exchanges, respectively. In the model of Ref.\ \cite{Glozman:1997fs,Glozman:1997ag}
this term is taken in the form
\begin{equation}
V_{F}({\vec x}_{\alpha})=\frac{g_{F}^{2}}{4 \pi} \frac{1}{12 m_{\beta} m_{\gamma}}
\left\{ \mu_{F}^{2} \frac{e^{-\mu_{F} x_{\alpha}}}{x_{\alpha}}  -\Lambda_{F}^{2} \frac{e^{-\Lambda_{F} x_{\alpha}}}{x_{\alpha}} \right\}~.
\end{equation}
The second Yukawa term is a smeared delta potential and its coupling constant $\Lambda_{F}$ is assumed to have a linear dependency on meson masses
\begin{equation}
\Lambda_{F}=\Lambda_{0}+\kappa \mu_{F}~,
\end{equation}
where $\Lambda_{0}$ and $\kappa$ are free parameters. In this model the masses of quarks and the mesons are fixed parameters, $m_{u}=m_{d}=340$ MeV, 
$m_{s}=500$ MeV, $\mu_{\pi}=139$ MeV, $\mu_{K}=494$ MeV, $\mu_{\eta}=547$ MeV, $\mu_{\eta'}=958$ MeV, and the meson octet-quark coupling constant was 
adopted as $g_{8}^{2}/ 4\pi=0.67$. 
The other parameters for the model were determined by fitting manually to the observed baryon spectra. Excellent agreement with experiment was found with
$V_{0}=416$ MeV and $C=2.33\ {\mathrm fm}^{-2}$, $\Lambda_{0}=2.86\ {\mathrm fm}^{-1}$, $\kappa=0.81$ and $(g_{0}/g_{8})^{2}=1.34$ 
\cite{Glozman:1997ag}.

It is natural to incorporate $V^{conf}$ into $v^{c}$ and $V^{\chi}$ into $v^{(s)}$. 
Then, to ensure that the poles of $G^{(c)}$ do not 
generate spurious solutions, we add a repulsive Gaussian term to $v^{(c)}$, which we subtract from
$v^{(s)}$
\begin{equation}
v^{(c)}  =V^{conf} + a_0 e^{-(r/r_0)^2}
\end{equation}
and 
\begin{equation}
v^{(s)} = V^{\chi}  - a_0 e^{-(r/r_0)^2}\ .
\end{equation}
 The parameters of the auxiliary potential have been taken as 
$a_0=3\;\mbox{fm}^{-1}$ and $r_0=1\;\mbox{fm}$. By this choice of the
parameter
values any bound states of $H^{(c)}$ are avoided below $\approx 2$ GeV. The
values
of $a_0$ and $r_0$ also influence the rate of convergence but not the
final (converged) results.

\section{Results}

In this approach we solve the three-quark problem in a discrete Hilbert space basis representation. 
To get reliable results we 
have to  achieve convergence in terms of basis states, i.e.\ in 
angular momentum channels, in terms of $N$ of  Eqs.\ (\ref{valpha-approx}), (\ref{ualpha-approx})
and (\ref{vsep}), and  in terms of $n$, which denotes the highest index in matrix continued fractions. 
Tables \ref{tab:1-1} and \ref{tab:1-2} show the convergence of the nucleon and excited nucleon state, 
respectively, in terms of $N$ and $n$.  We used 10 angular momentum channels as specified in Table \ref{tab:2}. We can see that 
$N=20$ and $n=60$ provides very accurate results. In calculating other results we kept these
basis sizes  fixed.

Table \ref{tab:2} shows the convergence of the nucleon and excited nucleon states in terms of angular momentum 
channels. The quantum numbers of the channels are selected to meet the angular momentum algebra and the 
symmetry condition (\ref{symcond}).
In Table \ref{tab:3} we present the results of our detailed calculations for various light baryon states. 
For comparison we also provide the results of a variational calculation 
\cite{Glozman:1997ag} and the results of our previous Faddeev calculations where the confinement were 
treated in an approximate way \cite{Papp:2000kp}.

\section{Summary}

In this work we presented an improvement over our previous semirelativistic Faddeev calculations of light baryons 
\cite{Papp:2000kp}. The main improvement is the exact treatment of the asymptotic confinement. For this purpose we 
calculated the channel Green's operator by a contour integral of two-body Green's operators. The two-body Green's
operators contain a linear confinement term, and they were evaluated with the help of a matrix continued fraction. Numerical
studies show some improvement over the previous solution method, but the difference is little, well within the range of experimental
errors. So, within the framework of Faddeev method, we have a good numerical solution of the semirelativistic three-body Hamiltonians of baryons.

\begin{table}[h]
\caption{Convergence of the nucleon (N(939)) and excited nucleon (N(1710)) mass in terms 
of $n$ of ,   the terms used in evaluating the matrix continued fraction. We used $N=24$ is the separable
expansion of the potential operators in the Faddeev equations.}
\label{tab:1-1}
\begin{tabular}{|c | c | c |c| }\hline\noalign{\smallskip}    
$N$ & $n$ & N(939) (MeV) & N(1710) (MeV)\\
\hline\noalign{\smallskip}   
24 & 30&  938.08  &  1775.85  \\
24 & 40&  938.06 &   1775.84\\
24 & 50&  938.05 &   1775.84  \\
24 & 60&  938.04 &   1775.84 \\
24 & 70&  938.04 &   1775.84 \\
24 & 80&  938.04 &   1775.83 \\
24 & 90&  938.04 &   1775.83 \\
\hline\noalign{\smallskip}   
\end{tabular}
\label{n939a}
\end{table}%

\begin{table}[h]
\caption{Convergence of the nucleon (N(939)) and excited nucleon (N(1710)) mass in terms 
of $N$,   the terms used in the separable
expansion of the potential operators in the Faddeev equations. The matrix continued fractions were evaluated up to $n=60$.}
\label{tab:1-2}
\begin{tabular}{|c | c | c | c|}
\hline\noalign{\smallskip}   
$N$ & $n$ & N(939) (MeV) & N(1710) (MeV)\\
\hline\noalign{\smallskip}  
16 & 60 & 938.16 & 1775.86 \\
17 & 60 & 938.14 & 1775.85 \\
18 & 60 & 938.12 & 1775.85 \\
19 & 60 & 938.11 & 1775.85 \\
20 & 60 & 938.10 & 1775.85 \\
21 & 60 & 938.08 & 1775.84 \\
22 & 60 & 938.07 & 1775.84 \\
23 & 60 & 938.06 & 1775.84 \\
24 & 60 & 938.05 & 1775.84  \\
\hline\noalign{\smallskip}  
\end{tabular}
\label{n939b}
\end{table}%

\begin{table}
\caption{Masses of the nucleon ground state and the first two positive-parity
excitations for the semirelativistic  GBE chiral quark model. 
The convergence with respect to including an increasing number of
angular-momentum states $l_x,l_y$ is demonstrated; $s$ and $t$ denote the
subsystem spin and isospin. In all cases, $N=20$ separable terms have been
included into the expansion. A comparison to the results obtained with the
previous calculations is given.}
\label{tab:2}
 \begin{tabular}{|c|cccc|c|c|c|}
\hline\noalign{\smallskip} 
\multicolumn{5}{|c|}{\mbox{Channels}}&\multicolumn{1}{|c|}{\mbox{Nucleon}}&
\multicolumn{1}{|c|}{$N^\ast$(1440)}&\multicolumn{1}{|c|}{$N^\ast$(1710)} \\
\hline\noalign{\smallskip}   
\# & $l_x$ & $l_y$ & $s$&$t$&\mbox{}& \mbox{}&\mbox{} \\ \hline\noalign{\smallskip}   
1   & 0 & 0  & 0 & 0 & 1615.6    & 1991.7     & 2294.2    \\
2   & 0 & 0  & 1 & 1 &  954.3     &  1470.8 &   1911.3  \\
3   & 1 & 1  & 1 & 0 &  952.8     & 1469.0   &  1907.3     \\
4   & 1 & 1  & 0 & 1 &  950.6     &   1465.7 &  1793.7    \\
5   & 2 & 2  & 0 & 0 &  946.6    &  1463.1  &  1792.7 \\
6   & 2 & 2  & 1 & 1 &  939.4    &   1459.0  & 1782.6    \\
7   & 3 & 3  & 1 & 0 &  939.3    &   1458.8  & 1780.0  \\
8   & 3 & 3  & 0 & 1 &  939.2    &  1458.6  &  1776.9  \\
9   & 4 & 4  & 0 & 0 &  938.7    &   1458.3 &  1776.9   \\
10 & 4 & 4  & 1 & 1 &  938.1     &  1457.8  & 1775.9 \\
\hline\noalign{\smallskip}    
\multicolumn{5}{|c|}{\mbox{ Ref.\ \cite{Papp:2000kp}  }}& 939 & 1459 &1775  \\ \hline\noalign{\smallskip}  
\multicolumn{5}{|c|}{\mbox{ Ref.\ \cite{Glozman:1997ag}  }}& 939 & 1459 &1776  \\ \hline\noalign{\smallskip}  
\end{tabular}
\end{table}

\begin{table}[h]
\caption{Masses of light baryons.}
\begin{tabular}{| l | c |c| c |c |}
\hline\noalign{\smallskip}
Baryon  & Ref.\ \cite{Papp:2000kp} & Ref.\ \cite{Glozman:1997ag} & This work &  Experiment  \\
\noalign{\smallskip}\hline\noalign{\smallskip}
N(939) &         938   & 939 & 938.0      &     938-940  \\
N(1440)            &    1458   & 1459 &  1457.8    &     1420-1470  \\
N(1520)           &    1517   & 1519 &  1517.0   &     1515-1525  \\
N(1675)           &    1644   & 1647 &  1644.4   &     1670-1680  \\
N(1710)            &   1776    & 1776 &    1775.8   &     1680-1740  \\
$\Delta(1232)$ &    1239   & 1240 &1238.8   &     1231-1233  \\
$\Delta(1600)$ &    1717   & 1718 &  1717.2    &     1550-1700  \\
$\Delta(1620)$ &    1639   & 1642 &  1638.9   &     1600-1660  \\
\noalign{\smallskip}\hline
\end{tabular}
\label{tab:3}
\end{table}

\bibliographystyle{spphys}
\bibliography{3quark}

\begin{thebibliography}{10}
\providecommand{\url}[1]{{#1}}
\providecommand{\urlprefix}{URL }
\expandafter\ifx\csname urlstyle\endcsname\relax
  \providecommand{\doi}[1]{DOI \discretionary{}{}{}#1}\else
  \providecommand{\doi}{DOI \discretionary{}{}{}\begingroup
  \urlstyle{rm}\Url}\fi

\bibitem{Friar:1980za}
J.L. Friar, B.F. Gibson, G.L. Payne, Phys. Rev. C \textbf{22}(1), 284 (1980).
\newblock \doi{10.1103/PhysRevC.22.284}

\bibitem{SilvestreBrac:1985ic}
B.~Silvestre-Brac, C.~Gignoux, Phys. Rev. D \textbf{32}(3), 743 (1985).
\newblock \doi{10.1103/PhysRevD.32.743}

\bibitem{Glozman:1996wq}
L.Y. Glozman, Z.~Papp, W.~Plessas, Phys. Lett. \textbf{B381}, 311 (1996).
\newblock \doi{10.1016/0370-2693(96)00610-7}

\bibitem{Papp:1998yt}
Z.~Papp, Few Body Systems \textbf{26}, 99 (1999)

\bibitem{Papp:2000kp}
Z.~Papp, A.~Krassnigg, W.~Plessas, Phys. Rev. C \textbf{62}(4), 044004 (2000).
\newblock \doi{10.1103/PhysRevC.62.044004}

\bibitem{Kelbert:2007}
E.~Kelbert, A.~Hyder, F.~Demir, Z.T. Hlousek, Z.~Papp, Journal of Physics A:
  Mathematical and Theoretical \textbf{40}(27), 7721 (2007).
\newblock \urlprefix\url{http://stacks.iop.org/1751-8121/40/7721}

\bibitem{Glozman:1997ag}
L.Y. Glozman, W.~Plessas, K.~Varga, R.F. Wagenbrunn, Phys. Rev. D
  \textbf{58}(9), 094030 (1998).
\newblock \doi{10.1103/PhysRevD.58.094030}

\bibitem{merkuriev1980three}
S.P. Merkuriev, Annals of Physics \textbf{130}, 395 (1980)

\bibitem{PhysRevLett.33.1350}
G.~Gignoux, A.~Laverne, S.P. Merkuriev, Phys. Rev. Lett. \textbf{33}(22), 1350
  (1974).
\newblock \doi{10.1103/PhysRevLett.33.1350}

\bibitem{PhysRevA.45.2723}
C.Y. Hu, A.A. Kvitsinsky, S.P. Merkuriev, Phys. Rev. A \textbf{45}(5), 2723
  (1992).
\newblock \doi{10.1103/PhysRevA.45.2723}

\bibitem{PhysRevA.63.062721}
Z.~Papp, C.Y. Hu, Z.T. Hlousek, B.~K\'onya, S.L. Yakovlev, Phys. Rev. A
  \textbf{63}(6), 062721 (2001).
\newblock \doi{10.1103/PhysRevA.63.062721}

\bibitem{Papp:1996yf}
Z.~Papp, W.~Plessas, Phys. Rev. C \textbf{54}(1), 50 (1996).
\newblock \doi{10.1103/PhysRevC.54.50}

\bibitem{Darai:2001yj}
B.~K\'{o}nya, G.~L\'{e}vai, Z.~Papp, Journal of Mathematical Physics
  \textbf{38}(9), 4832 (1997).
\newblock \doi{10.1063/1.532127}.
\newblock \urlprefix\url{http://link.aip.org/link/?JMP/38/4832/1}

\bibitem{kelbert2007green}
E.~Kelbert, A.~Hyder, F.~Demir, Z.~Hlousek, Z.~Papp, Journal of Physics A:
  Mathematical and Theoretical \textbf{40}, 7721 (2007)

\bibitem{Day:2009ax}
J.~Day, J.~McEwen, Z.~Papp, Few Body Systems \textbf{47}(1), 17 (2010).
\newblock \doi{10.1007/s00601-009-0063-2}

\bibitem{Glozman:1997fs}
L.Y. Glozman, Z.~Papp, W.~Plessas, K.~Varga, R.F. Wagenbrunn, Phys. Rev. C
  \textbf{57}(6), 3406 (1998).
\newblock \doi{10.1103/PhysRevC.57.3406}

\end{thebibliography}

\end{document}